\documentclass[12pt]{article}
\usepackage{makeidx}
\usepackage{a4,color,graphics,palatino}
\usepackage{amsmath}
\usepackage{amsfonts}
\usepackage{amssymb}
\usepackage{graphicx}
\usepackage{epsfig}
\usepackage[all]{xy}
\usepackage{fancyhdr}
\title{{\bf Magnetic and Electric Black Holes\\ in Arbitrary Dimension}}

\author{Adil Belhaj$^{1,2}$, Pablo D\'iaz$^1$ and Antonio Segu\'i$^1$\bigskip \\
$^{1}${\small Departamento de F\'{\i}sica Te\'{o}rica, Universidad de
Zaragoza}\\
{\small Pedro Cerbuna, 12. E-50009 Zaragoza, Spain}\medskip\\
$^2${\small Centre National de l`Energie, des Sciences et des Techniques Nucleaires},\\
{\small CNESTEN,
Cellule Sciences de la Matiere, Rabat, Morocco}\\
{\small and Groupement National de Physique des Hautes Energies},\\ 
{\small GNPHE,
Siege focal: FS, Rabat, Morocco}}

\date{}
\begin{document}
\maketitle

\begin{abstract}
In this work, we compare two different objects: electric black holes and magnetic black holes in arbitrary dimension. The comparison is made in terms of the corresponding moduli space and their  extremal geometries.  We treat parallelly the magnetic and the electric cases. 
Specifically, we   discuss  the gravitational solution of these spherically symmetric objects in the presence of  a positive cosmological constant. Then, we find the  bounded region of the moduli space  allowing the existence of black holes. After identifying it in both the electric and the magnetic case, we calculate the geometry that comes out between the horizons at the coalescence points. Although the electric and magnetic cases are both very different (only dual in four dimensions), gravity solutions seem to clear up most of the differences and lead to very similar geometries.
\end{abstract}

\newpage

\bigskip

\tableofcontents
\maketitle

\section{Introduction}
Reissner-Nordstrom black holes are static, spherically symmetric configurations which minimize the Maxwell-Einstein action. The solution in four dimensions was first found in~\cite{RN}. Every solution of this kind is completely defined by giving two parameters: the charge of the black hole Q and the mass M. The generalization to higher dimensional spacetimes with a cosmological constant was given by Tangherlini in~\cite{Tangherlini}. For certain range of parameters (see~\cite{cardoso} for a detailed description), the geometry of these objects present three horizons: Cauchy, black hole and cosmological. The thermodynamical properties of black holes permits those system to dynamically vary some parameters of the moduli space. For instance, the evaporation process may reduce the mass of a charge black hole to the point of coalescence at which the two inner horizons lead to a degenerate solution called the extreme black hole. In semiclassical relativity, extreme black holes are uncapable to emit radiation\footnote{See~\cite{Randall} for a recent discussion on the differences between semiclassical and string/Quantum Gravity counting of microstates for the computation of the entropy in extreme black holes.}. For this reason they are commonly known as cold black holes.

Magnetically charged objects have being considered since Dirac first claimed the theoretical existence of magnetic monopoles~\cite{D} in $U(1)$ electromagnetic theories in four dimensions. For  certain values of their parameters, magnetic monopoles can undergo a gravitational collapse and form black holes. They are magnetic black holes. Lubkin suggested that the magnetic charge of a monopole should be considered as a topological charge~\cite{lubkin}. Since then on, magnetic charges have been regarded as labels for the various topologies a field configuration can present. In particular, the identification magnetic charge/topology permits an easy generalization of the monopole concept to higher dimensions and different gauge groups. 

In~\cite{GP}, Ginsparg and Perry realized that some physical space between the horizons remains at the coalescence point. They studied the neutral Schwarzschild-de Sitter geometry. They actually showed that a Nariai geometry, which is the direct product $dS_2\times S^2$, came out in this process. 
The same technique has been profusely applied to some two-simple-horizons systems (electrically charged, magnetically charged, rotating black holes...) at the point where they coalesce and develop a degenerate horizon. The result is a collection of Nariai and anti-Nariai solutions, i.e. $dS_2\times S^{d-1}$ and $AdS_2\times S^{d-1}$ in $d+1$-spacetime dimension, whose radii relation is encoded in the details of each setup.   

The aim of this work is to compare two different objects: electric black holes and magnetic black holes in arbitrary  dimensions. The comparison is made in terms of the corresponding moduli space and the extremal near horizon geometries they present. We treat parallelly the magnetic and the electric cases. 
The magnetic object is not determined without referring it to a concrete gauge symmetry group of a theory. Even the extension of the Dirac monopole is not unique. Although we discuss deeper on this topic in the next section, for the moment, let us just say that in this paper we deal with the orthogonal extension (Yang-type) of the Dirac-Yang series. Specifically, we use the gravitational solution of these spherically symmetric objects with the addition of a positive cosmological constant found in~\cite{townsend}. Then, applying an analogous analysis to~\cite{cardoso} for the magnetic case, we consider the  bounded region of the moduli space which allows the existence of black holes. After identifying it in both the electric and the magnetic case, we compute  the geometry that comes out between the horizons at the coalescence points. There we find the following three cases:
\begin{itemize}
\item The coalescence of the cosmological and black hole horizon which leads to a generalized Nariai solution.\\
\item The coalescence of the Cauchy and the black hole horizon which produces an anti-Nariai solution.\\
\item The triple coalescence of horizons which makes a product geometry $\mathcal{M}^{1,1}\times S^{d-1}$. 
\end{itemize}
 The paper is organised as follows. We start with a section of mathematical preliminaries in order to fix notation and establish the basic mathematical framework. Section \ref{sec:PSEMBH} is devoted to finding the moduli space for magnetic black holes and compare it to the electric case. The coalescence geometries are all studied along section \ref{sec:EGMCYtMBH}. In particular, we propose a different change of coordinates to the one used in~\cite{cardoso} and we discuss the consistency of our election. Finally, in the  conclusions section, we sum up the main results and discuss the similarities and subtle differences between the electric and the magnetic system.  

\section{Preliminaries}

\begin{figure}
\begin{center}
\includegraphics{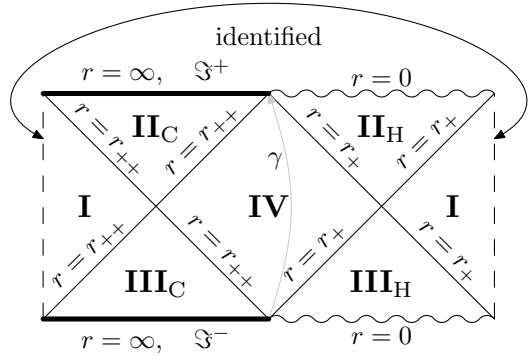}
\caption{{\small
Carter-Penrose diagram of maximally extended Schwarzschild-de Sitter black hole.}}
\label{fig:SdS}
\end{center}
\end{figure}

De Sitter-Reissner-Nordstr\"om (electrically charged) black holes in higher dimensions are static, spherically symmetric configurations which minimize the Einstein-Hilbert action
\begin{equation}\label{eq:einsteinaction}
S=\int dx^{d+1} \sqrt{-g}\big(\frac{1}{16\pi}(R-2\lambda)+\frac{1}{4}F^2\big),
\end{equation}
in a $d+1$ spacetime. The field strength $F$ is a closed to form. Indeed, it can always be locally written as $F=dA$, where $A$ is the potential 1-form.
Varying action (\ref{eq:einsteinaction}) with respect to the metric tensor $g_{\mu\nu}$ leads to the well-known Tangherlini metric~\cite{Tangherlini}:
\begin{equation}\label{eq:RNdSmetric}
ds^2=-\Delta_e(r)dt^2+\Delta_e^{-1}(r)dr^2+r^2d\Omega^2_{d-1},
\end{equation}
where
\begin{equation}\label{eq:RNdS}
\Delta_e(r)=1-\frac{2m}{r^{d-2}}+\frac{Q^2}{r^{2(d-2)}}-\frac{r^2}{R^2}.
\end{equation}  
The subscript $e$ stands  for ``electric''. The cosmological constant $\Lambda$ is a function of the de Sitter radius $R$ and $d$ is the dimension $d$  where we have the relation  $R=\sqrt{\frac{d(d-1)}{2\Lambda}}$.\\

After the classical Dirac monopole, the first significant example of monopoles in higher dimensional pure\footnote{If we allow a scalar (Higgs) field to enter the theory then regular solutions can be found in any dimension. See~\cite{Pol74,tH74} for the  4-dimensional case and~\cite{Tch2000,Tc,Tch2008} for examples in arbitrary dimension}  into the  gauge theories is the Yang construction~\cite{Y}. They are point-like objects of a $SU(2)$-gauge theory in  6-dimensional spacetime (see~\cite{BDS1} and the references therein for a recent string realization of the Yang monopole). It is a natural step to consider $G$ groups for theories in $(d+1)$-dimensional spacetime which are a generalization of both Dirac and Yang objects. There are at least to ways of thinking of a such a generalization~\cite{ours}. {\it Yang-type} or $SO(2n)$ monopoles are orthogonal extensions of this series. The ones we are going to treat in this paper. These constructions were first found in~\cite{Nep84}, widely studied in~\cite{Tch2000,Tc,Tch2008} and recently reviewed in~\cite{townsend,ours,thesis}. They are topologically nontrivial solutions of a $SO(2n)$-gauge theory in ($2n+2$)-dimensions. For $n=1$ we obtain the Dirac monopole, after the identification of the isomorphic groups $SO(2)$ and $U(1)$. For $n=2$ the Yang-type object is {\it not} the Yang monopole but the extended-Yang monopole, which belongs to a $SO(4)$ gauge theory, instead of the $SU(2)$ theory of the Yang monopole. The fact that $SO(4)\cong \frac{SU(2)\times SU(2)}{\mathcal{Z}_2}$ triggered the claim that they were actually two copies of the Yang monopole~\cite{townsend}. There is another possible continuation of the Dirac-Yang series. They are the more recent $SU(2^{n-1})$-monopoles introduced by G. Meng~\cite{meng} in $2n+2$ dimensions, where the Dirac and Yang constructions can be recover for $n=1,2$, respectively.\\
In general, static and spherically symmetric solitons may be classified for any dimension $d$ and gauge group $G$ (see~\cite{ours,thesis}).  Given an arbitrary gauge group $G$ and a dimension $d$, point-like static solitons may not exist. Their existence basically depends on the possibility of nontrivial homomorphisms $\lambda:SO(d-1)\to G$ (see~\cite{ours}). Indeed there is a one-to-one map between solitonic configurations of this type and $\lambda$-homomorphisms up to isomorphism~\cite{wang}.\\

Point-like static magnetic objects in $(d+1)$-dimensional spacetime are specific solitonic solutions of a $G$-gauge theory, where $G$ is the Lie group of symmetry of the theory. They are $A$ configurations\footnote{For a detailed description of the potential and field strengths see~\cite{townsend,thesis}} which minimize the Yang-Mills-Hilbert-Einstein action
\begin{equation}\label{EHYMaction}
S = \int dx^{d+1} \sqrt{-g}\,  \left[ {1\over 16\pi } \left(R -2\Lambda\right)
- {1\over2\gamma^2}\, Tr |F|^2  \right],
\end{equation} 
that is, configurations which fulfill the Einstein equations:
\begin{equation}\label{eq:EA}
G_{\mu\nu}=8\pi  T_{\mu\nu}-g_{\mu\nu}\Lambda,
\end{equation}
where
\begin{equation}\label{eq:EMtensor}
T_{\mu\nu}=\gamma^{-2}\bigg[\mathrm{tr} (F_{\mu}^{\phantom{\mu}p} F_{\nu p})-\frac{1}{4}g_{\mu\nu}\mathrm{tr} (F_{pq}F^{pq})\bigg]
\end{equation}
is the energy momentum tensor of the YM field strength. The traces are taken in the colour index which labels the basis of $\mathfrak{g}$ and $\gamma$ is the YM coupling constant. 

Yang-type solutions of (\ref{EHYMaction}) were found in~\cite{townsend}.  Finding general solutions for (\ref{eq:EA}) is a highly complicated problem. 
However, being point-like makes the solution spherically symmetric. Imposing spherical symmetry simplifies the task enormously. According to this, in Schwarzschild like coordinates the ansatz will be a spherically symmetric $(2k+2)$-dimensional space whose line element reads
\begin{equation}\label{eq:metric}
ds^{2}=-\Delta dt^{2}+\Delta^{-1} dr^{2}+r^{2}d\Omega^{2}_{2k}.
\end{equation}
The ansatz (\ref{eq:metric}) is consistent with (\ref{eq:EA}) and (\ref{eq:EMtensor}) for
\begin{equation}
\Delta(r)=1-\frac{2M(r)}{r^{2k-1}}.
\end{equation}
Now, $M(r)$ can be integrated  to give
\begin{equation}\label{eq:mag}
\Delta_m(r)=1-\frac{2m}{r^{2k-1}}-\frac{\mu^{2}}{r^{2}}-\frac{r^{2}}{R^{2}},
\end{equation}
where $R=\sqrt{\frac{k(2k+1)}{\Lambda}}$ is the de Sitter radius, $\mu^2$ is proportional to $\frac{1}{2k-3}$ and measures the magnetic charge of the monopole, $m$ comes up as a constant of integration with dimensions of mass. The subscript $m$ stands for ``magnetic''. The moduli space of the magnetic solution is then $\{m,\mu,\Lambda,d\}$. \\

\begin{figure}
\begin{center}
\includegraphics{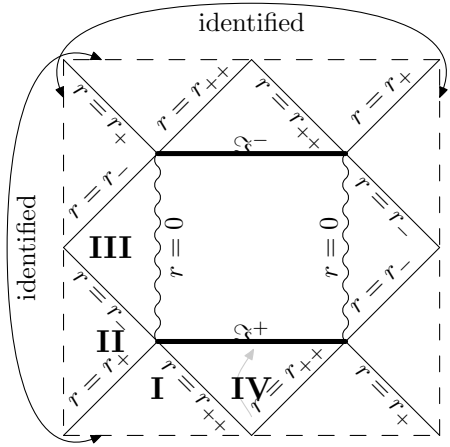}
\caption{{\small
Carter-Penrose diagram of the maximally extended de Sitter-Reissner-Nordstr\"om solution.}}
\label{fig:dSRN}
\end{center}
\end{figure}

   The only difference between (\ref{eq:RNdS}) and (\ref{eq:mag}) is found
    in the term involving the charge $Q$.  The sign of this term differs
     for $d>3$, for $d\leq 3$ as in 4-dimensional spacetime, $\mu^2$ is negative~\cite{townsend}.
   But the essential change is the order of that term, which is $r^{-2(d-2)}$
   for the electrical case as Tangherlini~\cite{Tangherlini} first found,
    and $r^{-2}$ for the magnetic system\footnote{Note that in the magnetic
     case this term does not depend on the dimensionality of spacetime.}.
    
\begin{figure}[h]
\begin{center}
\includegraphics{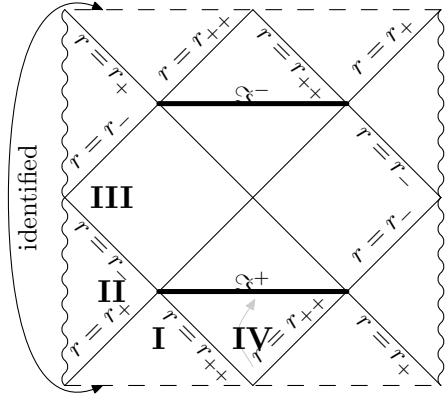}
\caption{{\small
Carter-Penrose diagram of the maximally extended de Sitter-Reissner-Nordstr\"om extremal solution.}}
\label{fig:dSRNextremal}
\end{center}
\end{figure}

As in the electric setup, the geometry driven by (\ref{eq:mag}) is asymptotically
de Sitter, since the
 behavior of $\Delta_m(r)$ for long distances $r>>R$ is governed by the last term. It means, in particular, that there always exists a cosmological horizon whose radial position is labeled by $r_{++}$. From the point of view of causality, more interesting results are obtained when black holes also enter in the discussion,
   what occurs for a certain range of parameters. So in spite of the differences above-mentioned, it can be shown that for a certain region
     of the moduli space there appears three horizons: Cauchy, Black hole and
     Cosmological horizons, which is an analogous  situation to the one
      studied in~\cite{DS} for the case $m<0$ . Let us note that the magnetic object
       for $m>0$, in which no Cauchy horizon appears, has no analogy in the electrical side.

If we forget the topological arguments given all along~\cite{DS}
about $\mu^2$
 being quantized which forbid us to finely tune it{\footnote{It is known that Yang-type monopoles are not the only spherically symmetric solutions. There is actually a family of solutions parmetrized by an essential function $w(r)$. In a recent conversation with Eugen Radu, he suggested that this function could actually make the work to obtain continous values for charge $\mu^2$. We will investigate on that issue in the future.}, and
  consider $\mu^2$ a continuous variable as $Q^2$ for the electrical case,
  we can draw a picture $m(\mu^2)$ to show the regions where non-extreme
  black holes appears in the theory for a fix spacetime dimension $d+1$
  and a fix (positive) cosmological constant implicit in $R^2$. This allows
   us to open a comparison between the electric and the magnetic setups in
   higher than four dimensions. The permitted region for non-extreme black
    holes is always bound by the lines where two horizons coalesce. For the
    electric object as well as for the magnetic one (if $m<0$) there are
    three horizons, labeled by $r_-$, $r_+$ and $r_{++}$ in order of increasing radial position. They lead to two lines in the diagram which correspond to:
\begin{itemize}
\item The coalescence of Cauchy and the black hole horizon. It leads to extreme
 Reissner-Nordstrom (RN) black holes\footnote{Also called cold black holes.}.
\item The coalescence of black holes and cosmological horizon which produces
Charged Nariai (CN) black holes.
\end{itemize}
The analysis  made in ~\cite{DS} tackles only the second process
since
 the Cauchy horizon is not always present in the magnetic setup. However, a similar procedure we followed in~\cite{DS} and inspired
   in the work of~\cite{GP} to obtain Nariai solutions can also be done for
    the coalescence of inner horizons, as we see later. Before going into the geometries, our first task is identifying the boundary lines which
         enclose the parameter region $\{m,\mu^2\}$ for non-extreme black
         holes in  the magnetic case. Let
         us see how it goes.

\section{Parameter Space of Electric and Magnetic Black Holes}\label{sec:PSEMBH}
In~\cite{DS} we found the  region of the moduli space $\{m,\mu,\lambda,d\}$ which allows the existence of magnetic black holes, and the values of the parameters which saturate the inequality $r_+\leq r_{++}$  and lead to the coalescence solution. In this section we aim at treating this study in a more compact manner as done in~\cite{cardoso} for the electric case, and depict a final magnetic diagram were the region of existence of black holes is bounded by the line in the moduli space for the three type of  coalescence solutions. We will see the differences and similarities on both electric and magnetic systems and interpret them.   The notation we will take is the same as used in~\cite{DS}, although the radial coordinate for the coalescence point in each case will be called $\rho$ instead of $r_c$, for the sake of compactness. \\

\begin{figure}[h]
\begin{center}
\includegraphics{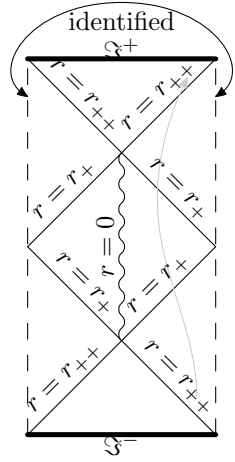}
\caption{{\small
Carter-Penrose diagram of the maximally extended charged Nariai black hole.}}
\label{fig:CNBH}
\end{center}
\end{figure}

\subsection{The coalescence of two horizons}
The coalescence of two Killing horizons takes place whenever the metric
 component $\Delta_m(r)$, displayed in (\ref{eq:mag}), has two roots.
 Then $\Delta_m(r)$ can be written as
\begin{equation}\label{eq:deltatwo}
\Delta_m(r)=(r-\rho)^2\frac{1}{r^2}\bigg[1-\frac{1}{R^2}(r^2+h(r))\bigg],
\end{equation}
where $r=\rho$ is the degenerate horizon. For function $h(r)$ we will
 take the ansatz
\begin{equation}\label{eq:h}
h(r)=a+br+\frac{c_1}{r}+\frac{c_2}{r^2}+\dots +\frac{c_{d-4}}{r^{d-4}},
\end{equation}
where $a,b,c_1,\dots,c_{d-4}$ are constants (functions of $\rho$, actually)
 that are determined through the matching order by order between
 (\ref{eq:mag})
  and (\ref{eq:deltatwo}). Function $h(r)$ turns out to be
\begin{equation}
h(r)=3\rho^2+2\rho r+\frac{4\rho^3-2\rho R^2}{r}+\frac{\mu^2 R^2+5\rho^4-3\rho^2
 R^2}{r^2}+\sum_{i=1}^{d-6}\frac{(i)}{\rho^{i+1} r^{d-(i+3)}}mR^2.
\end{equation}
Besides, the matching also permit to write parameters $m$ and $\mu^2$
 as functions of $\rho$, they read
\begin{eqnarray}
\mu^2(\rho)&=&\frac{\rho^2}{d-4}\bigg[d-2-d\frac{\rho^2}{R^2}\bigg] \label{eq:muderho}\\
m(\rho)&=&\frac{2}{d-4}\rho^{d-2}\bigg[2\frac{\rho^2}{R^2}-1\bigg] \label{eq:mderho}.
\end{eqnarray}
An analogous analysis for the electric case would involve (\ref{eq:RNdS}) and lead to equations~\cite{cardoso}
\begin{eqnarray}
Q^2(\rho)&=&\rho^{2(d-2)}\bigg[1-\frac{d}{d-2}\frac{\rho^2}{R^2}\bigg] \label{eq:Qderho}\\
m(\rho)&=&2\rho^{d-2}\bigg[1-\frac{d-1}{d-2}\frac{\rho^2}{R^2}\bigg] \label{eq:emederho}.
\end{eqnarray}
It is clear from equations (\ref{eq:muderho}) and (\ref{eq:mderho})
 that the dimension of spacetime must be different from five.
 As discussed in~\cite{DS} and more carefully in~\cite{ours}, the magnetic monopoles
  we are working on are even dimensional, so $d=4$ is never
  going to be the case. For $d=3$, $\mu^2$ becomes negative~\cite{townsend},
  as can easily be seen in (\ref{eq:muderho}), and the extreme
  Reissner-Nordstr\"om (electric or magnetic) solution is recovered. The cases
  we are interested in concern spacetime dimension $d\geq 2k+1$,
   with $k=2,3,\dots$, which will be assumed in the following. Now,
   $\mu^2$ being positive imposes a maximum on $\rho$:
\begin{equation}\label{eq:rhomax}
\rho_{max}=\sqrt{\frac{d-2}{d}}R.
\end{equation}
To obtain a relation $m(\mu^2)$ from (\ref{eq:muderho}) and (\ref{eq:mderho})
is an easy task. We must invert equation (\ref{eq:muderho}) to
 get\footnote{It turns out that independently of the value of
 $\mu^2$, the relation $\rho_{-}<\rho_{+}<\rho_{max}$ holds.}
 two functions $\rho^2_{+}(\mu^2)$ and $\rho^2_{-}(\mu^2)$.
 Then we should plug them into (\ref{eq:mderho}). The result
 is a straightforward calculation, although the ugliness of
 the relation has stopped us from including it here. Instead,
  we will just say
\begin{equation}
m_{\pm}\propto \pm \mu^{\frac{d}{2}}.
\end{equation}

\subsection{Coalescence of three horizons or Ultracold Black Hole}

As for the electric example, for a concrete value of $m$ and $\mu^2$
there occurs a triple degeneracy of horizons. For the de
 Sitter-Reissner-Nordstrom solution it was called ``ultracold black hole'',
  name that we shall keep for the magnetic black holes.  The triple degeneracy
  for magnetic objects should happen for $m<0$ which as discussed before,
  is the only region where three horizons are expected to exist. We will check
   this fact for consistency, although it is easily seen in {\bf figure \ref{fig:mag}},
   where the sharp point of triple degeneracy is indicated.
\begin{figure}[h]
\begin{center}
\includegraphics{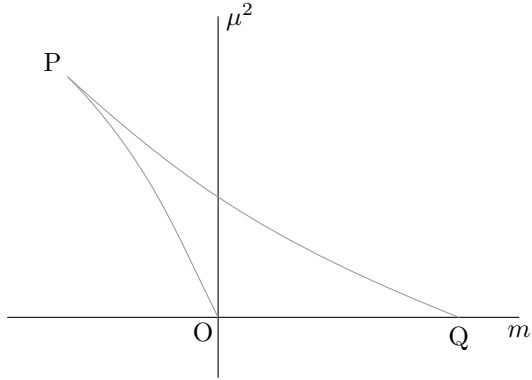}
\caption{{\small
The region bound by OPQ represent the pairs $\{m,\mu^2\}$
for which the magnetic black hole is not extreme. The line
OP is the parameter region for extremal (cold) black holes;
 PQ line corresponds to generalized Nariai solutions and the OQ
  segment stands for SdS geometries. The point Q represents
  the neutral Nariai black hole and the point P, the Ultracold
  black hole.}}

\label{fig:mag}
\end{center}
\end{figure}
\begin{figure}[h]
\begin{center}
\includegraphics{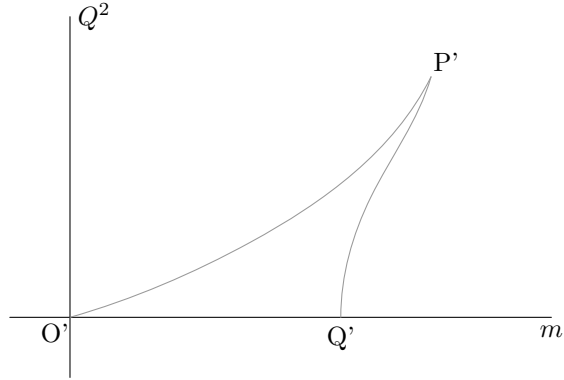}
\caption{{\small The
 region bound by O'P'Q' represent the pairs $\{m,Q^2\}$ for
 which the electric black hole is not extreme. The line Q'P'
  is the parameter region for extremal (cold) black holes; P'Q'
  line corresponds to generalized Nariai solutions and the O'Q' segment stands for SdS geometries. The point Q' is represents the neutral Nariai black hole and the point P' the Ultracold black hole.}}
\label{fig:elec}
\end{center}
\end{figure}

The specific shape we have  given  to $\Delta_m(r)$ for the
coalescence
 of two horizons in (\ref{eq:deltatwo}) will guide us to find the
  very point where the triple coalescence takes place.
  The condition that must be imposed is no other than
\begin{equation}\label{eq:thcoalescence}
1-\frac{1}{R^2}(\rho^2+h(\rho))=0.
\end{equation}
Equation (\ref{eq:thcoalescence}) is quadratic and easily solved,
 it leads to
\begin{equation}\label{eq:rhocoal}
\rho_{mc}=\sqrt{\frac{d-2}{2d}}R=\frac{1}{\sqrt{2}}\rho_{max},
\end{equation}
where $mc$ stands for the triplet coalescence point in the magnetic case.
Now, substituting $\rho_{mc}$ into (\ref{eq:muderho}) and
(\ref{eq:mderho}), we get
\begin{eqnarray}
\mu^2(\rho_{mc})&=&\frac{(d-2)^2}{4d(d-4)}R^2, \nonumber \\
m(\rho_{mc})&=&-\frac{4}{d(d-4)}\rho_{mc}^{d-2},
\end{eqnarray}
respectively.  Note that $m(\rho_{mc})$ is manifestly negative for $d>4$.

The triple degeneracy point $P$ in the magnetic case is similar to $P'$ in
 the electric diagram ({\bf figure \ref{fig:elec}}), for this reason we keep
  the name. In the electric case, the coalescence point, the mass and the charge  which were found
  in~\cite{cardoso} are given  by
\begin{eqnarray}\label{eq:coalelec}
\rho_{ec}&=&\frac{d-2}{\sqrt{d(d-1)}}R, \nonumber \\
Q^2(\rho_{ec})&=&\frac{1}{d-1}\rho_{ec}^{2(d-2)}, \nonumber \\
m(\rho_{ec})&=&\frac{4}{d}\rho_{ec}^{d-2},
\end{eqnarray}
where subscript $ec$ labels the triple coalescence point for the electric case.
There are some differences between the electric and magnetic diagrams
which are worth explaining. The line $O'P'$ consists of the set of pairs
 $(m,Q)$ where extreme black holes (Reissner-Nordstrom) are formed.
 The two horizons which coalesce all along this line are the Cauchy and
 the Black hole horizons. In the diagram, it is shown that even in the case
  that a given value of $m$ permits the coalescence of the other two exterior
   horizons, the one which lies on $O'P'$ has maximum charge. This is the reason
    why they are also called cold black holes. The line $P'Q'$, which closes the
     figure, correspond to Nariai-like solutions. These degenerate black holes
      are extremal in the sense of ``packing'' a given charge with the maximum mass.
       Now, given the symmetry of the diagrams, one might be wrongly tempted to
        associate line $PQ$ to cold black holes and $OP$ to Nariai kinds.
        It is actually the other way round. To see this, let us go back to
         the electric side and seek the relevant magnitude that makes the distinction. It turns out to be the ratio
\begin{equation}
\lambda_e=\frac{Q^2}{m}R^{d-2},
\end{equation}
where the de Sitter radius $R$ has been included to make
 $\lambda_e$ dimensionless. Now, let us choose any value
  of $m$, if there are two possible extreme black holes for it,
   then, the Nariai kind will be the one for which $\lambda_e$ is smaller. So, we conclude:
\begin{equation}
\lambda_e^{Nariai}\leq \lambda_e^{Cold}.
\end{equation}
The inequality getting saturated at the triple degeneracy point $P'$.
Moreover, using equations (\ref{eq:Qderho}) and (\ref{eq:emederho})
 and evaluating $\lambda_e$ along the extremal path $O'P'Q'$ reveals
  that it has a maximum at $P'$. So, in the electric case, we may
  characterize the ``ultracold black holes'' by having a maximum $\lambda_e$.
   Analogously, we will choose the dimensionless magnitude
\begin{equation}
\lambda_m=\frac{\mu^2}{m}R^{d-4},
\end{equation}
and impose
\begin{equation}
\lambda_m^{Nariai}\leq \lambda_m^{Cold}.
\end{equation}
Here enters the fact that in the magnetic case $m$ is negative for
 the cases with three horizons. This reverses the inequality for
  absolute values and, particularly, makes line $OP$ of pairs $(m,\mu)$ be the ``cold black hole''
  parameter boundary. The line $PQ$ is the Nariai boundary, consequently. Again, using equations
   (\ref{eq:muderho}) and (\ref{eq:mderho}) and evaluating $\lambda_m$ along the pairs which
   lie on the extreme path reveals the existence of a maximum at $P$. So, we can conclude
    that, as in the electric case, the triple degeneracy point corresponds to magnetic
     black holes with maximal rate $\lambda_m$, which justifies the name of ``ultracold black holes''.

\section{Extremal geometries of magnetically\\ charged (Yang-type) black holes}\label{sec:EGMCYtMBH}
In this section we apply the procedure found by Ginsparg and Perry~\cite{GP} to describe
 the geometries for the extreme magnetically charged black holes as they come from non-extremality.
  In~\cite[appendix A]{DS} it was proved that there is always some physical space left between two simple coalescent horizons. It means that 
  there is always a geometry at the coalescence point which cannot be described with our initial set of coordinates. Indeed, there is always a change of coordinates that makes this geometry manifest.
We will start this section by reviewing the Generalized Nariai Solutions, already found in~\cite{DS}.
  Then we calculate the necessary different limits which leads to dS-Bertotti-Robinson,
  $\Lambda=0$ Bertotti-Robinson and Nariai-Bertotti-Robinson solutions. {\bf A comparison
  with the results found in~\cite{cardoso}} for the electric case will be kept always in mind.
\subsection{Generalized Nariai solutions}
In~\cite{DS} we found a ``wise change of coordinates'' which will be regular and,
 consequently suitable to describe the geometry between the horizons, at
 the coalescence point. We will now make use of the notation and the mechanism describe
  in the previous section in order to present all the results in a compact form.
   Generating Nariai kind of solutions from the near-Nariai black holes can be
    done as follows. Near the coalescence point,  the black hole and the cosmological
     horizons are located at $r_+=\rho-\epsilon$ and $r_-=\rho+\epsilon$ respectively.
      With this parametrization coalescence takes place as $\epsilon \to 0$.
      Function $\Delta(r)$ in (\ref{eq:mag})  can be rewritten in the form
\begin{equation}\label{eq:deltaGN}
\Delta(r)=-A^{N}(r)\big(r-\rho(1+\epsilon)\big)\big(r-\rho(1-\epsilon)\big),
\end{equation}
where
\begin{equation}\label{eq:Ader}
A^{N}(r)=-\frac{1}{r^2}\bigg[1-\frac{1}{R^2}(r^2+h(r))\bigg],
\end{equation}
according to (\ref{eq:deltatwo}). Thus, the degenerate horizon of
the black hole is placed at $r=\rho$. The change of coordinates we
need to perform to obtain the ``inner'' geometry is
\begin{eqnarray}\label{eq:coordinatesnariai}
t&=&\frac{\tau}{\epsilon A(\rho)} \nonumber \\
r&=&\rho(1+\epsilon\cos\chi).
\end{eqnarray}
The next step is to apply (\ref{eq:coordinatesnariai}) to (\ref{eq:deltaGN})
and take the limit $\epsilon \to 0$. The result can be used to compute the
 quantities $-\Delta(r)dt^2$, $\Delta^{-1}(r)dr^2$ and $r^2$ in the new
  coordinates $(\tau,\chi)$. In this way,  the generalized Nariai line element
  comes out from (\ref{eq:RNdSmetric}) as
\begin{equation}\label{eq:Nariailineelement}
ds_{N}^2=\frac{1}{A^N(\rho)}(-\sin^2\chi d\tau^2+d\chi^2)+\frac{1}{\rho^2}d\Omega^2_{d-1}.
\end{equation}
This is a generalized Nariai geometry in the sense of being the direct product
 $dS_2 \times S^{d-1}$, that is, a $(1+1)$-dimensional dS spacetime with radius
  $\frac{1}{A^{N}(\rho)}$ and a $(d-1)$-sphere of radius $\frac{1}{\rho^2}$.
  The factor $A(\rho)$ can be calculated in the electric and the magnetic case.
   They turn out to be
\begin{eqnarray}\label{eq:Aes}
A_{e}^{N}(\rho)&=&(d-2)^2\frac{1}{\rho^2}-d(d-1)\frac{1}{R^2} \nonumber \\
A_m^{N}(\rho)&=&(d-2)\frac{1}{\rho^2}-2d\frac{1}{R^2}.
\end{eqnarray}
\subsection{Generalized cold solution}
The generalized cold (de Sitter Bertotti-Robinson) black hole solution
is generated using an analogous technique as in the previous section.
This time the coalescence takes place between the two inner  (the Cauchy
 and the black hole) horizons. Now, the $g_{00}$ component of  metric (\ref{eq:RNdSmetric})
  will be written as
\begin{equation}\label{eq:deltaGdSBR}
\Delta(r)=A^{Cold}(r)\big(r-\rho(1+\epsilon)\big)\big(r-\rho(1-\epsilon)\big),
\end{equation}
where a parametrization of the location of the horizons $r_-=\rho(1-\epsilon)$
and $r_+=\rho(1+\epsilon)$ have already been used. A degenerate horizon geometry
 is again obtained as $\epsilon \to 0$. A suitable change of coordinates,
\begin{eqnarray}\label{eq:coordinatesdSBR}
t&=&\frac{\tau}{\epsilon A^{Cold}(\rho)} \nonumber \\
r&=&\rho(1+\epsilon\cosh\chi),
\end{eqnarray}
applied to (\ref{eq:deltaGdSBR}), followed by the limit $\epsilon \to 0$ and
 inserted in  (\ref{eq:RNdSmetric}), permits us to find the gravitational field
 of the cold black hole solution
\begin{equation}\label{eq:dSRBlineelement}
ds_{Cold}^2=\frac{1}{A^{Cold}(\rho)}(-\sinh^2\chi d\tau^2+d\chi^2)+\frac{1}{\rho^2}d\Omega^2_{d-1}.
\end{equation}
This is a generalized dS Bertotti-Robinson geometry in the sense of being
the direct product $AdS_2 \times S^{d-1}$, that is, a $(1+1)$-dimensional AdS spacetime with radius $\frac{1}{A^{Cold}(\rho)}$ and a $(d-1)$-sphere of radius $\frac{1}{\rho^2}$. The factor $A^{Cold}(\rho)$ can be calculated in the electric and the magnetic case. It is easy to see that

\begin{eqnarray}\label{eq:AesdSBR}
A_{e}^{Cold}(\rho)&=& -A_{e}^{N}(\rho) \nonumber \\
A_m^{Cold}(\rho)&=& -A_{m}^{N}(\rho).
\end{eqnarray}
The cold flat (Bertotti-Robinson~\cite{Bertotti,Robinson}) solution is obtained by the limit $\Lambda \to 0$ ($R \to \infty$) in the cold black hole geometry. The result is again a geometrical product $AdS_2 \times S^{d-1}$ with the radius of the AdS space, $1/A^{(BR)}(\rho)$, given by
\begin{eqnarray}\label{eq:AesBR}
A_{e}^{(BR)}&=& -(d-2)^2\frac{1}{\rho^2} \nonumber \\
A_m^{(BR)}(\rho)&=& -(d-2)\frac{1}{\rho^2}.
\end{eqnarray}
\subsection{Generalized ultracold solution}\label{GUCS}
Let us generate a generalized ultracold (Nariai-Bertotti-Robinson)
solution from the near triple degeneracy point. Recall that $\rho_{ec,mc}$ stands
 for the triple coalescence point of the electric and the magnetic ultracold black hole,
  respectively. The change of coordinates we need to perform is 
\begin{eqnarray}\label{eq:coordinatesUC}
t_{e,m}&=&\frac{\tau}{a_{e,m}\epsilon^{3/2}} \nonumber \\
r_{e,m}&=&\rho_{ce,cm}(1+\epsilon\cos\sqrt{b_{e,m}\epsilon^{1/2} \chi}),
\end{eqnarray}
 where
\begin{eqnarray}
a_e&=&(d-2)\sqrt{\frac{2}{3}} \nonumber \\
a_m&=&\sqrt{\frac{2(d-2)}{3}}
\end{eqnarray}
and
\begin{eqnarray}
b_e&=&\frac{4}{R}\sqrt{\frac{d(d-1)}{3}} \nonumber \\
b_m&=&\frac{4}{R}\sqrt{\frac{d}{3}}.
\end{eqnarray}
Substituting (\ref{eq:coordinatesUC}) into (\ref{eq:RNdS}) and (\ref{eq:mag})
one can easily calculate the quantities $-\Delta(r)dt^2$, $\Delta^{-1}(r)dr^2$
 and $r^2$ in the new coordinates at the coalescence point. The ultracold geometries
  for both the electric and the magnetic case are
\begin{eqnarray}\label{eq:ple}
ds^2_e&=&-d\tau^2+d\chi^2+\rho_{ec}^2 d\Omega^2_{d-1} \\
ds^2_m&=&-d\tau^2+d\chi^2+\rho_{mc}^2 d\Omega^2_{d-1}.
\end{eqnarray}
In both cases, the geometry turns out to be $\mathcal{M}^{1,1}\times
S^{d-1}$. This $d+1$ dimensional geometries are causally equivalent to 2-dimensional Minkowski
spacetime.  For $d=3$, this kind of solution was discussed
in~\cite{plebanskian}. In this sense it
 would be fair to call them ``generalized plebanski-hacyan'' solutions.\\
 
 \begin{figure}[h]
\begin{center}
\includegraphics{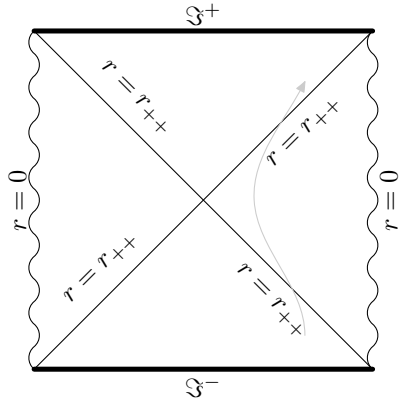}
\caption{{\small
Carter-Penrose diagram of the maximally extended ultra cold black hole.}}
\label{fig:UltraCold}
\end{center}
\end{figure}

There is a discrepancy with~\cite{cardoso} about the change of coordinates (\ref{eq:coordinatesUC}) 
   we have performed in both the electric and the magnetic case for ultracold solutions. They proposed $r\to\rho (1+\epsilon\cos\sqrt{2A^{UC}\epsilon}\chi)$ and $t=\tau/(A^{UC}\epsilon^2)$. As we argue below, their change of
   coordinates is adapted to a particular path for the coalescence process whereas (\ref{eq:coordinatesUC}) is valid for  
   more general ones. There is, however, no differences (it would otherwise have been worried) in the
   geometry obtained at the triple coalescence point.  
   
In order to generate a generalized ultracold (Nariai-Bertotti-Robinson)
 solution from the near triple degeneracy point,  the authors of~\cite{cardoso}
  wrote  the function $\Delta(r)$ as
\begin{equation}\label{eq:UCcardoso}
\Delta(r)=-A^{UC}(r)\big(r-\rho_{ec,mc}(1+\epsilon)\big)^2\big(r-\rho_{ec,mc}(1-\epsilon)\big),
\end{equation}
where, again, $\rho_{ce,cm}$ stands respectively for the triple coalescence
point of the electric
 and the magnetic ultracold black hole.
Equation (\ref{eq:UCcardoso}) assumes that the two exterior horizons
 have already coalesce and the Cauchy horizon comes to them at the
  triple degeneracy point. In  { \bf figure \ref{fig:elec}}, it accounts for reaching $P'$ along
   the path $Q'P'$. The transformation
\begin{eqnarray}\label{eq:cardosotrans}
r&\to&\rho_e(1+\epsilon\cos(\sqrt{2 b\epsilon}\chi)) \nonumber \\
t&\to&\frac{\tau}{a\epsilon^2},
\end{eqnarray}
together with taking the coalescence-value parameters $\{\rho,m,Q^2\}$
  as displayed in (\ref{eq:coalelec}) into (\ref{eq:UCcardoso})
   along this path, produces leading terms of the order $\epsilon^4$,
    which cancel out the fourth order divergence of $dt^2$ and $dr^2$
    in the new coordinates. In this way, the coalescence geometry is
    nonsingular. 
     Things would have not been that nice and regular had we chosen the inner two horizons
     to coalesce first and the   outer one to join the triplet. That is,
     if we had considered
\begin{equation}\label{eq:UCcardoso2}
\Delta(r)=-A^{UC}(r)\big(r-\rho_{ec,mc}(1+\epsilon)\big)\big(r-\rho_{ec,mc}(1-\epsilon)\big)^2,
\end{equation}
which accounts as an approach to $P'$ by the path $O'P'$, then the
change of coordinates (\ref{eq:cardosotrans}) would lead to a
singular geometry. The correct transformation would have needed to
involve a $\cosh$ function instead. It seems that the
transformation we should perform depends on the path to be taken.
Does it make any sense?  The paths may seem capriciously chosen so
far. A general $\Delta(r)$ function with three horizons to form a
triple degeneracy point is (in coordinates $\{t,r\}$) singular to
leading order $\epsilon^3$ if
 the approaches are linear in $\epsilon$. After the change of coordinates,
  the square 1-forms $dt^2$ and $dr^2$ must be singular up to order
   $\epsilon^{-3}$ and $\epsilon^3$, respectively. This is all achieved
    by a change of coordinates of the type displayed in (\ref{eq:coordinatesUC}).\\
     However, unless a given point has different limits depending on the path is taken (what happens for singular points),
     there is in principle no problem in adjusting the change of
      coordinates to the coalescence path, for changes of coordinates are
       just mathematical artifacts which help to blow up an apparent singularity.
        The final finite geometry must be the same whatever path is taken.
         Precisely, it is the cartesian product of (1+1)-Minkowsky and $(d-1)$-spheres.
         
\section{Conclusion}
The coupling to gravity of two objects in arbitrary dimension $d$ coming 
from completely different theories have been studied
in this paper. The electric black hole, which is a perturbative solution
of an $U(1)$-gauge theory, has been put together with a solitonic configuration
of a $SO(d-1)$-gauge theory: the magnetic (Yang-type) black hole. The line elements
of both are similar except for the term involving the charge. Nevertheless, this 
does not make much difference in the geometrical analysis. The bounded
region of the moduli space for nonextreme black holes in the magnetic setup 
has been calculated and depicted. We have found a complete analogy with the electric case.
As far as the geometries are concerned, the differences are minor, and subject always to
constant factors which depend mainly on the dimension. Again we encountered generalized 
Nariai, generalized anti-Nariai and $\mathcal{M}^{1,1}\times S^{d-1}$  metrics after the three possible
coalescence processes. The causal regions are the same in the magnetic case, fact that 
enables us to use the same Carter-Penrose conformal diagrams to describe them an depicted in figures \ref{fig:SdS}, \ref{fig:dSRN}, \ref{fig:dSRNextremal}, \ref{fig:CNBH} and \ref{fig:UltraCold}. \\

We find interesting that such completely different systems (only dual in 4-dimensional spacetime~\cite{HR})
share similar gravitational and causal properties. In this case, it is not due to an intrinsic
duality of the gauge objects but to the action of gravity which clears out the main differences.\\

A recently written paper~\cite{Randall} has shown that there are actually
two ways of understanding the limit of coalescence in the electric 4-dimensional case
(it is straightforward to extend this result to arbitrary dimension). One leads to 
extremal black hole solutions, and the other one gives   the compactified solutions. Both  are locally 
identical but different at a global scale. They suggest that this two different limits count for the discrepancy
of the entropy computation between semiclassical gravity and string theory. Specifically,
they say that both computations do not agree because they are not referred to the same geometry. 
We strongly believe that
magnetic black holes enjoy a similar status. Indeed, there is a plan for a future work
in which we will perform the entropy computations for the extreme (cold) magnetic black hole.\\

\noindent\textbf{Acknowledgments.} This work has been supported by CICYT (grant
FPA-2006-02315) and DGIID-DGA (grant 2007-E24/2) and by the department of theoretical physics of the University of Zaragoza. 
The authors are indebted to Vitor Cardoso, Andrew DeBenedictis, Oscar Dias, Jose Lemos,
Francisco Navarro-Lerida, Istvan Racz  and Eugen Radu for their critical 
reading and the valuable comments on the manuscript.

\end{document}